\documentclass[12pt,a4paper]{article}
\usepackage{latexsym}
\usepackage{hyperref}
\newlength{\altura}
\newcommand{\old}[1]{\settoheight{\altura}{$#1$}{\stackrel{{\rm o}}{#1}}\rule[0pt]{0pt}{\altura}}
\newcommand{\be}{\begin{equation}}
\newcommand{\ee}{\end{equation}}

\begin{document}
\title{Conservation of Energy--Momentum in Teleparallel Gravity.}
\author{M. Hermida de La Rica\\ 
Facultad de Inform\'atica.
Universidad Polit\'ecnica de Madrid.\\
Universidad Complutense de Madrid.\\
University of Ngozi.}

\maketitle
\begin{abstract}
In a well-known paper \cite{Andrade2000} V.C. de Andrade,
L. C. T. Guillen and J.G. Pereira defined a conserved gauge current
$hj_a^{\;\;\rho}$, however they stated that: ``{\sl This is, we
believe the farthest one can go in the direction of a tensorial
definition for the energy and momentum of the gravitational field. The
lack of local Lorentz covariance can be considered as the teleparallel
manifestation of the pseudotensor character of the gravitational
energy--momentum density in general relativity\ldots}''. Well, we
believe that they stopped just less than an inch before giving such a
tensorial definition, and furthermore that the resulting
energy--momentum tensor has zero trace and can be made symmetric, as a
matter of fact it is just $j^{(\nu\rho)}$, and together with the
energy--momentum tensor of material fields it obeys a natural
conservation equation for teleparallel manifolds. Some important
consequences are obtained specially in the last section concerning the
possibility of explaining the acceleration of the universe expansion
without any need of a cosmological constant.
\end{abstract}

\section{The energy--momentum tensor}
We follow most of the conventions and notations of references
\cite{Andrade2000} and \cite{Andrade1997}. However one difference is
that we stick to consider the lagragian density of the gravitational
field to be proportional to $+\old{R}$ (as it is usually accepted)
instead of being proportional to $-\old{R}$ as is done in
\cite{Andrade2000}. Any further difference will be
made it explicit. Anyway let us at least remember that greek indices
are used for the coordinate holonomic quantities and latin indices are
used for non--holonomic ones.

One way of viewing the kind of Riemannian spaces in which teleparallel
theories are formulated is by taking as departure point the hypothesis
that physics somehow establishes a canonical, smooth, path-independent
isomorphism between the tangent spaces of any two points of the
manifold and hence that we may take some orthonormal, but otherwise
arbitrary reference basis, which we will call $\vec{u}_a$, and refer
all vectors at all points to that basis (of course, we identify
$\vec{u}_a$ with its preimage at any given point). Parallel transport
can then be introduced as meaning to transport keeping constant
components with respect to this basis, then Cartan covariant
derivative is just the variation with respect to this reference basis
expressed, for example, in terms of the coordinate
basis. Mathematically, this has the consequence of accepting only
parallelizable manifolds as physically meaningful, this is nothing
more than just a topological condition on the sort of Riemannian
spaces we deal with. In particular the coordinate vectors
$\partial_\mu(x)$, no matter which sets of coordinates $x$ we take,
should be expressible in terms of the reference basis:
\begin{equation} 
\partial_\mu(x) = h^{a}{}_{\mu}(x)\vec{u}_a
\end{equation}
Let us also remember that the relationship between the Levi-Civita
covariant derivative $\old{\nabla}$ due to the symmetric riemannian
connection and the ``Cartan'' covariant derivative $\nabla$ of the
Weitzenb\"ock connection is given by the difference between the
Christoffel symbols of both covariant derivatives:
\begin{equation}
\Gamma^\rho{}_{\mu\nu}=\old{\Gamma}^\rho{}_{\mu\nu}+ K^\rho{}_{\mu\nu}
\end{equation}
 Where $K^\nu{}_{\rho\nu}$ is the contorsion tensor given by:
\begin{equation}
K^\rho{}_{\mu\nu}= \frac{1}{2}\left(g^{\rho\alpha}\left[T_{\mu\alpha\nu} + T_{\nu\alpha\mu}\right] - T^\rho{}_{\mu\nu}\right)
\end{equation}
Now, as a first point in the discussion we must clarify what can be considered
a conservation equation for the energy--momentum tensor within teleparallel
theory.  Suppose we had some (symmetric) energy--momentum tensor
$S_{\mu}^{\;\;\nu}$ and let $w^\mu$ be the components of some vector which is
Cartan covariant constant. Such vectors do exist because any linear constant
combination of vectors of the reference basis $\vec{u}_a$ is Cartan
covariant constant. $S_{\mu}^{\;\;\nu} w^\mu$ represents the flow of the
component of energy--momentum in the direction of the four-vector $\vec{w}$, so
$\old{\nabla}_\nu(S_{\mu}^{\;\;\nu} w^\mu)=0$ expresses the conservation of
such $\vec{w}$ component. However we can write this as:
\begin{eqnarray}\nonumber
0 & = &\old{\nabla}_\nu(S_{\mu}^{\;\;\nu} w^\mu)=\nabla_\nu(S_{\mu}^{\;\;\nu}
w^\mu) - S_{\mu}^{\;\;\rho} w^\mu K^\nu{}_{\rho\nu}\\
& = & w^\mu\left(\nabla_\nu S_{\mu}^{\;\;\nu} - S_{\mu}^{\;\;\rho} T^\nu{}_{\nu\rho}\right)
\end{eqnarray}
If we want conservation of energy--momentum in all directions, then it
must hold:
\begin{equation}
\nabla_\nu S_{\mu}^{\;\;\nu} - S_{\mu}^{\;\;\rho} T^\nu{}_{\nu\rho}=0\label{conserved1}
\end{equation}
This is not the same as the condition $\old{\nabla}_\nu
S_{\mu}^{\;\;\nu}=0$. If we substitute the Cartan covariant derivative
by its classical counterpart, we reach another expression:
\begin{equation}
\old{\nabla}_\nu S_{\mu}^{\;\;\nu} - S_{\sigma}^{\;\;\rho} K^\sigma{}_{\mu\rho}=0
\end{equation}
Of course, these equations reduce to the zero divergence condition in
Minkowski spaces, however the important point is that (for second rank
tensors in teleparallel spaces) they represent the correct
generalization of the zero divergence condition of Minkowski space.

As it is well known Einstein's tensor $G^{\mu\nu}$ verifies
$\old{\nabla}_\nu G^{\mu\nu}=0$, so from Einstein's equation we do not
get a conservation equation for the energy--momentum tensor of matter
alone.  For the energy--momentum tensor $\cal T $ of matter, accepting
it to be a symmetric tensor, we rather get:
\begin{equation}
0 = \old{\nabla}_\nu {\cal T}_{\mu}^{\;\;\nu}\quad \Longrightarrow\quad 
 \left(\nabla_\nu {\cal T}_{\mu}^{\;\;\nu} -
{\cal T}_{\mu}^{\;\;\rho} T^\nu{}_{\nu\rho}\right)  + {\cal
  T}_{\sigma}^{\;\;\nu}T^\sigma{}_{\nu\mu}  = 0 \label{divergmat}
\end{equation}
Comparing it with (\ref{conserved1}) we see it is not quite exactly
the same, a further term is present. Let us remember that we are
dealing with a lagrangian density $\Lambda$ for the gravitational
field which in teleparallel theories might in general be written as:
\begin{equation}
 \Lambda= \kappa_g(a_1\Lambda_1 + a_2\Lambda_2 + a_3\Lambda_3)
\end{equation} 
where:
\begin{eqnarray}
\Lambda_1  & = &  g^{\lambda\mu}T^\alpha{}_{\alpha \lambda}T^\beta{}_{\beta \mu}\qquad
\Lambda_2  =  g^{\lambda \mu}T^\alpha{}_{\beta \lambda}T^\beta{}_{\alpha \mu} \\
\Lambda_3 & = &  T^{\rho \beta \mu}T_{\rho \beta \mu}\qquad
\kappa_g  =  \frac{c^4}{16\pi G}
\end{eqnarray}

The ``default'' values: $a_1=1$, $a_2=-1/2$, $a_3=-1/4$, produce a
lagrangian density equivalent to general relativity, meaning that for
those values the difference between $\old{R}$ and $(a_1\Lambda_1 +
a_2\Lambda_2 + a_3\Lambda_3)$ is just a total divergence. As a matter
of fact, adding
$2g^{\alpha\beta}\old{\nabla}_{\alpha}T^{\lambda}{}_{\beta\lambda}$
one obtains $\old{R}$. However, for the moment, we are not going to
fix the value of those coefficients, we want to keep open the
possibility of choosing other values for them.

One way of obtaining the gravitational energy--momentum tensor is to
directly reproduce the reasoning which in classical mechanics leads to
the conservation of energy and write (we follow the standard work
\cite{Land73}, which as a matter of fact is just the first step of
applying Noether's method considering the translational invariance of
the lagrangian, see also \cite{Leclerc2006}, section 2):
\begin{equation}
\frac{\partial (\Lambda h)}{\partial x^\mu}
= \frac{\partial (\Lambda h)}{\partial h^a_{\;\;\nu}}\frac{\partial
  h^a_{\;\;\nu}}{\partial x^\mu} + \frac{\partial (\Lambda h)}{\partial
  h^a_{\;\;\nu ,\mu}}\frac{\partial h^a_{\;\;\nu ,\mu}}{\partial x^\mu}
\end{equation}
The field equations can be written as:
\begin{equation}
\frac{1}{h}h^a_{\;\;\sigma}\left[\frac{\partial (\Lambda h)}{\partial h^a_{\;\;\nu}} -
\frac{\partial}{\partial x^\gamma}\left( h \frac{\partial
\Lambda}{\partial h^a_{\;\;\nu ,\gamma}}\right)\right] + {\cal T}_{\sigma}^{\;\;\nu} = 0\label{Euler}
\end{equation}
And taking them into account one inmediately is led to:
\begin{equation}
{\frac{1}{h}}\frac{\partial}{\partial
  x^\gamma}\left[h\left(h^a_{\;\;\nu ,\mu}\frac{\partial \Lambda}{\partial
    h^a_{\;\;\nu ,\gamma}} - \Lambda\delta^\gamma_\mu\right)\right] -
{\cal T}_{\sigma}^{\;\;\nu}\Gamma^{\sigma}_{\;\;\nu \mu}= 0
\end{equation}
One would like to identify the term within the round brackets in last
equation with the energy--momentum tensor, but it is not a tensor, so
the idea is to decompose it into tensorial and non--tensorial terms,
hence we write previous equation as:
\begin{equation}
\frac{1}{h}\frac{\partial}{\partial x^\gamma}\left[h(Q_{\mu}^{\;\;\gamma} + N_{\mu}^{\;\;\gamma})\right] -
{\cal T}_{\sigma}^{\;\;\nu}\Gamma^\sigma_{\;\;\nu \mu} = 0 \label{entangled}
\end{equation}
Where $Q_{\mu}^{\;\;\gamma}$ is the tensorial part and
$N_{\mu}^{\;\;\gamma}$ is a non--tensorial term. Expressing the
partial derivative of the $Q$ tensor as a Cartan covariant derivative
one arrives inmediately to the following equation:
\begin{equation}
 0 = \nabla_\gamma Q_{\mu}^{\;\;\gamma}  -  Q_{\mu}^{\;\;\gamma} T^\nu{}_{\nu\gamma} +
  \Gamma^\nu{}_{\nu \gamma}N_{\mu}^{\;\;\gamma} 
+  Q_{\sigma}^{\;\;\gamma}\Gamma^\sigma{}_{\mu \gamma} +  \frac{\partial N_{\mu}^{\;\;\gamma}}{\partial x^\gamma} -
{\cal T}_{\sigma}^{\;\;\nu}\Gamma^\sigma{}_{\nu \mu}\label{nontensorial}
\end{equation}
Now the problem is to eliminate the non--tensorial terms of this
equation. Of course, it must be possible to do so, because you cannot
have an equalty between entities which transform in different ways
(you may put the first two terms at one side and the other terms at the
other side). The needed calculations for eliminating, or transforming,
those non--tensorial terms are a bit cumbersome, but in the appendix
it is indicated how it can be shown that by taking
$Q_{\mu}^{\;\;\gamma}$ as:
\begin{eqnarray}\nonumber
Q_{\mu}^{\;\;\gamma} & = &
\kappa_g\left\{a_1\left[2\left(g^{\nu\gamma}T^\alpha{}_{\alpha
    \nu}T^\beta{}_{\beta \mu} - g^{u\beta}T^\alpha{}_{\alpha
    \nu}T^\gamma{}_{\beta \mu}\right)\right]\right. \\ \nonumber
& + &  a_2\left[2\left( g^{\nu\gamma}T^\alpha{}_{\beta
     \nu}T^\beta{}_{\alpha \mu} - g^{\nu\alpha}T^\gamma{}_{\beta
     \nu}T^\beta{}_{\alpha \mu}\right)\right]\\ 
& + &  a_3\left[4g^{\beta\gamma}g^{\alpha\nu}g_{\rho \tau}T^\rho{}_{\alpha
    \beta}T^\tau{}_{\nu \mu}\right] 
   -  \left.\Big[ a_1\Lambda_1 +a_2\Lambda_2 + a_3\Lambda_3\Big]\delta^\gamma_\mu\right\}
\end{eqnarray}
one arrives to the conclusion that
\begin{equation}
 \Gamma^\nu{}_{\nu \gamma}N_{\mu}^{\;\;\gamma} 
+  Q_{\sigma}^{\;\;\gamma}\Gamma^\sigma{}_{\mu \gamma} +  \frac{\partial N_{\mu}^{\;\;\gamma}}{\partial x^\gamma} =\Gamma^\sigma{}_{\mu \nu}{\cal
  T}_{\sigma}^{\;\;\nu}
\end{equation}
The expression for $Q_{\mu}^{\;\;\gamma}$ might seem strange at first sight, but it happens to be exactly $-j_{\mu}^{\;\;\gamma}$:
\begin{equation}
\frac{1}{h}h^a_{\;\;\mu}\frac{\partial (\Lambda h)}{\partial h^a_{\;\;\gamma}} = j_{\mu}^{\;\;\gamma} = -Q_{\mu}^{\;\;\gamma} 
\end{equation}
Hence we are led to the result:
\begin{equation}
\nabla_\gamma j_{\mu}^{\;\;\gamma} - j_{\mu}^{\;\;\gamma} T^\nu{}_{\nu\gamma} + {\cal T}_{\sigma}^{\;\;\nu} T^\sigma{}_{\mu \nu} = 0 \label{diverggrav}
\end{equation}
So taking into account both equations (\ref{diverggrav}) and (\ref{divergmat})
one gets:
\begin{equation}
\nabla_\gamma({\cal T}_{\mu}^{\;\;\gamma} + j_{\mu}^{\;\;\gamma}) - ({\cal T}_{\mu}^{\;\;\gamma} + j_{\mu}^{\;\;\gamma})T^\nu{}_{\gamma\nu} = 0
\end{equation}
Which has exactly the form of equation (\ref{conserved1}) and so it
can be interpreted as just expressing the conservation of total
energy--momentum: the energy--momentum of the material fields ${\cal
T}_{\mu}^{\;\;\gamma}$ plus the energy--momentum of the gravitational
field $j_{\mu}^{\;\;\gamma}$. It also clarifies the meaning of the
term ${\cal T}_{\sigma}^{\;\;\nu}T^\sigma{}_{\mu \nu}$. This term
specifies the energy--momentum interchange between the gravitational
field and the material one. It is the term which prevents
energy--momentum of the gravitational field or energy--momentum of the
material field from being conserved separately by themselves. The
interpretation of $j_{\mu}^{\;\;\gamma}$ as the correct
energy--momentum tensor for the gravitational field can be further
underlined if one rewrites the field equations as:
\begin{equation}
\frac{1}{h}h^a_{\;\;\sigma}\frac{\partial}{\partial x^\gamma}\left( h \frac{\partial
\Lambda}{\partial h^a_{\;\;\nu ,\gamma}}\right)  = j_\sigma^{\;\;\nu} + {\cal T}_{\sigma}^{\;\;\nu}
\end{equation}
and compares it with the equations for the electromagnetic field in
Minkowski space written as:
\begin{equation}
\frac{\partial}{\partial x^\gamma}\left( \frac{\partial
\Lambda_e}{\partial A_{\nu,\gamma}}\right) = -\frac{J^\nu}{c}
\end{equation}
Informally, it is usually accepted that this last equation says that
currents are the sources of the electromagnetic field. Then, in the
same sense, the previous equation might be interpreted as saying that
energy--momentum of both the material fields and the gravitational
field is the source of the gravitational field.

However, this is not the final word about the tensor we need, because
$j_\sigma^{\;\;\nu}$ has the uncomfortable characteristic that in
general it is not symmetric. If we accept the default values for the
coefficients $a_1, a_2, a_3$ then there is an easy way out of the
problem: Einstein's equations are a total of ten equations, however
they do not completely determine the metric tensor. Diffeomorphisms
comprise the gauge freedom in general relativity (see \cite{Wald84}
pg. 438): any two solutions which are related by a diffeomorphism
represent the same physical solution. Now, if instead of considering
the metric tensor as the final solution of a gravitational problem, we
ask a complete solution of such a problem to be given by the
specification of the sixteen functions $h^a_{\;\;\sigma}$ which give
the coordinate basis vectors in terms of the arbitrary constant
reference basis, then we have some further freedom, because once the
metric tensor is given, we may have several ``square roots''
$h^a_{\;\;\sigma}$ which produce the same metric tensor: There are
sixteen arbitrary $h^a_{\;\;\sigma}$ functions and only ten
independent conditions imposed by the metric tensor.  We have lots of
``gauge freedom'', so let us use part of that freedom to decree that
the antisymmetric part of the energy--momentum tensor
$j_\sigma^{\;\;\nu}$ should be zero. This ``gauge condition'' amounts
just to a set of six equations, because in a four dimensional space an
antisymmetric second rank tensor has only six independent components.
So, although it is a crude way of counting degrees of freedom, we have
increased by six the number of unknown functions when substituting the
metric tensor as solution by the $h^a_{\;\;\sigma}$, but we have also
added six additional equations, so we expect not to have changed the
``gauge freedom''. The gauge condition can be written as:
\begin{equation}
j^{[\gamma\nu]} = 0 = \left( g^{\mu\gamma}T^\nu{}_{\beta \mu}-
g^{\mu\nu}T^\gamma{}_{\beta
  \mu}\right)g^{\lambda\beta}T^\alpha{}_{\alpha \lambda} -
\frac{1}{2}\left( g^{\mu\gamma}T^\nu{}_{\beta \lambda}-
g^{\mu\nu}T^\gamma{}_{\beta
  \lambda}\right)T^{\beta\lambda}{}_{\mu}
\end{equation}
Of course it is a covariant condition: true in one coordinate system
means true in all, so we are not limiting the set of coordinates in
which the theory is formulated: we are not imposing conditions on the
sort of diffeomorphisms which might be used. As a matter of fact, it
has previously been argued that teleparallel theories may have too
much gauge freedom (see \cite{Kopczynski}, \cite{Nester}) and that
they suffer from a problem of non--predictability of torsion. Although
a formal proof should be investigated, we clearly expect this gauge
condition to fix such problems. At least, the introduction of this
condition invalidates the reasoning supporting such assertions,
because clearly these additional six equations have not been taken
into account when studying the predictability of torsion. Furthermore:
being an algebraic condition on the torsion tensor, not every boundary
condition is acceptable, because the boundary condition must also obey
the gauge condition.

Accepting such a gauge condition, the energy--momentum of the gravitational
field turns out to be symmetric, which just means that
$j^{\gamma\nu}=j^{(\gamma\nu)}$. And furthermore it is inmediate to
check that it has zero trace. Needless to say it is a perfectly
covariant local definition of energy--momentum for the gravitational
field.

There is one further point which merits some comment. Teleparallel
theories have some degree of freedom in the way the coefficients are
chosen. However it is only for the case in which we obtain the
teleparallel equivalent to general relativity when the resulting
equations are symmetric (we obtain Einstein's tensor). So it is only
in this case in which we have the freedom to impose that the
energy--momentum tensor of the gravitational field should be symmetric
(as there is no other possibility in empty space). So this condition
eliminates the rest of possibilities for the coefficients.

We may even generalize: if we restrict teleparallel theories to just
this case, as it has been said before, we obtain symmetric equations
(Einstein's tensor), so what would happen if the energy--momentum
tensor of a material field were not symmetric?. The most natural
answer in this hypothetical case would be to change the gauge
condition so that the antisymmetric part of the energy--momentum
tensor of the gravitational field just cancels the antisymmetric part
of the energy--momentum tensor of the material fields, and only the
symmetric part plays a role in Einstein's equation.

\section{The energy content of homogeneous, isotropic universes.}
\label{TheEnergyContent}
The energy content of homogeneous, isotropic universes has already
been computed in other papers (see \cite{Vargas03}, for
example). However, although our results are quite similar, there are
some points which should be noted: previous works have used
pseudo--tensors and of course have never taken into account the gauge
condition. Usually they need to integrate over a space section. Given
that we are dealing with a isotropic, homogeneous universe, if we were
really working with a true gravitational energy density, one would
expect the energy density to be homogeneous, and if the total energy
turns out to be zero, it is quite unintuitive that it is not zero at
every point.  Of course, one answer is that usually authors are not
working with the true energy density of the gravitational field.

To calculate the energy content of the three cases of null, positive
and negative curvature, we must first find ``square roots'' of the
respective metrics. Not every square root is acceptable, we must also
impose the gauge condition: the energy--momentum tensor derived from
them should be symmetric. However for these problems we must not only
obtain symmetrical tensors, but also diagonal, otherwise we would have
some preferred direction in space. It must be noted that most
calculations in this section have been done with Maple$^\copyright$\
9.5 running on Ubuntu Linux.

The easiest case is the one of zero curvature. We use a ``cartesian''
coordinate system with coordinates $ct, x, y, z$, and we postulate the
following matrix of gravitational potential vectors (the role played
by the coordinate vectors is similar to that of vector potentials):
\begin{equation} 
h^a_{\;\;\alpha} =\, {\rm diag}\left( 1, a(t), a(t), a(t) \right) 
\end{equation}
Using this potentials, the metric is just the very well known diagonal
metric of flat space $g_{\nu \mu} = {\rm diag}(1,
-a^2(t),-a^2(t),-a^2(t))$. The energy--impulse tensor for such a space
is:
\begin{equation} 
j^0_0  =   -6\kappa_g\left( \frac{\dot{a}(t)}{a(t)}\right)^2\qquad
j^1_1=j^2_2=j^3_3  =   2\kappa_g\left( \frac{\dot{a}(t)}{a(t)}\right)^2
\end{equation}
Where the dot signals ordinary differentiation with respect to $ct$. The first
point which deserves attention is that energy density is negative, so it seems
that there is at least a known field whose energy density takes negative
values. It is also purelly ``kinetical'' in this case: it is proportional to
the square of the speed at which $a(t)$ changes, and the minus sign tells us
that absortion of (positive) energy will decrease this speed.

Let us consider first the case of a dust-filled universe. We know that
for such a case $a(t) = C_d t^{2/3}$, so the gravitational energy
density is proportional to $-t^{-2}$ which, when multiplied by
$a^3\propto t^2$ to take into account the increase in volume, just
gives constant energy: dust does not contribute to any variation of
energy of the gravitational field, it does not interchange energy with
the gravitational field.

Consider now the case of a universe filled with just radiation, being
$\rho$ its energy density. It is well-known that in such a case the
solution for $a(t)$ is of the form $a(t)=C_r t^{1/2}$. So the
gravitational energy density is also proportional to $-t^{-2}$ (it is the
square of a logarithmic derivative, so no matter the exponent it will
be proportional to $-t^{-2}$), which when multiplied by $a^3(t)$, to
take into account the increase of volume, gives the result that energy
of gravitational field changes as $-t^{-1/2}$, which is an increase
and which is just the rate needed to compensate the rate at which
energy of radiation decreases: $\rho a^4$ is constant, so $\rho a^3$
decreases as $a^{-1}\propto t^{-1/2}$. The absortion of positive
energy from radiation just decreases the rate at which universe
expands. In the dust-filled case, the decrease in speed is just to
compensate the increase in volume, so that total energy is the
same. As energy of light is absorved by the gravitational field, its
``kinetic'' energy increases (it decreases its ``speed''). A radiation
dominated universe expands at a slower rate ($t^{1/2}$) than a dust
filled one ($t^{2/3}$): absortion of energy decreases its speed.

Even more clear, for the flat universe Einstein's equation can be
used to calculate the energy--momentum tensor of matter:
\begin{equation} 
{\cal T}^0_0  =   6\kappa_g\left(\frac{\dot{a}}{a}\right)^2\qquad
{\cal T}^1_1 =  {\cal T}^2_2 = {\cal T}^3_3 =  2\kappa_g\left(\frac{\dot{a}^2+2\ddot{a}a}{a^2}\right)
\end{equation}
Looking at the first term we see that the energy density of the matter fields
is just the same (but positive) as the energy density of the gravitational
field, so that total energy density is zero. This is a fact which also happens
in the case of positive curvature.

For positive curvature universe the way to get an acceptable square root of
the metric is to remember that $S^3$ is parallelizable, so it is easy to get
three ortonormal vectors which are tangent to it and then take them as the
spatial part of the reference basis at each point. Of course, by an acceptable
square root we mean one which renders a symmetric gravitational
energy--momentum tensor. A diagonal square root, as it is usually taken, does
not. The previous idea can be summed up by saying that we postulate an
$h^a_{\;\;\alpha}$ matrix given by:

\begin{equation}{\scriptsize
\left( \begin{array}{llll}
1 & 0 & 0 & 0 \\
0 & a(t)c(\Theta) & -a(t)s(\Psi)c(\Psi)s(\Theta) & -a(t)s^2(\Psi)s^2(\Theta) \\
0 & a(t)s(\Theta)c(\Phi) & a(t)s(\Psi)\left(s(\Psi)s(\Phi)+ c(\Psi)c(\Theta)c(\Phi)\right)&  a(t)s(\Psi)s(\Theta)\left(s(\Psi)c(\Theta)c(\Phi) - c(\Psi)s(\Phi)\right) \\
0 & a(t)s(\Theta)s(\Phi) & a(t)s(\Psi)\left(c(\Psi)c(\Theta)s(\Phi) - s(\Psi)c(\Phi)\right) & a(t)s(\Psi)s(\Theta)\left(c(\Psi)c(\Phi) + s(\Psi)c(\Theta)s(\Phi)\right) \\
\end{array} \right)} 
\end{equation}

Where $s\equiv\sin$ and $c\equiv\cos$. This matrix leads to the spherical
metric:
\begin{equation} 
g_{\nu \mu} =  {\rm diag}(1,
 -a^2(t),-a^2(t)s^2(\Psi), -a^2(t)s^2(\Psi)s^2(\Theta))
\end{equation}
The energy--momentum tensor of matter is given by:
\begin{equation} 
{\cal T}^0_0   =  6\kappa_g\frac{\dot{a}^2 +1}{a^2}\qquad
 {\cal T}^1_1  = 
{\cal T}^2_2={\cal T}^3_3 =  2\kappa_g\frac{2\ddot{a}a + \dot{a}^2 + 1}{a^2}
\end{equation}
The gravitational energy--momentum tensor is given by:
\begin{equation}
j^0_0  =  -6\kappa_g\frac{\dot{a}^2 +1}{a^2}\qquad
 j^1_1=j^2_2=j^3_3  =  2\kappa_g\frac{\dot{a}^2 +1}{a^2}
\end{equation}
So we get into the same situation in which the total energy is
zero. 

It is a bit more difficult to find an acceptable square root for the
negative curvature case. As a matter of fact, in the acceptable
solution which has been found, the potential vectors cannot be so
neatly separated into spatial and temporal parts, however they do lead
to the correct metric and to a diagonal energy--momentum tensor. The
solution found for matrix $h^a_{\;\;\alpha}$ is:

\begin{equation}{\footnotesize 
\left( \begin{array}{llll}
ch(\Psi) & a(t)sh(\Psi) & 0 & 0 \\
sh(\Psi)c(\Theta) & a(t)ch(\Psi)c(\Theta) & -a(t)sh(\Psi)s(\Theta) & 0 \\ 
sh(\Psi)s(\Theta)c(\Phi) & a(t)ch(\Psi)s(\Theta)c(\Phi)& a(t)sh(\Psi)c(\Theta)c(\Phi) & -a(t)sh(\Psi)s(\Theta)s(\Phi) \\ 
sh(\Psi)s(\Theta)s(\Phi) & a(t)ch(\Psi)s(\Theta)s(\Phi) & a(t)sh(\Psi)c(\Theta)s(\Phi) & a(t)sh(\Psi)s(\Theta)c(\Phi)
\end{array} \right)} 
\end{equation}

Where $ch\equiv \cosh$ and $sh\equiv \sinh$. It is inmediate to check that
the hyperbolic metric is obtained with these potentials:
\begin{equation} 
g_{\nu \mu}  =  {\rm diag}(1, -a^2(t), -a^2(t)\sinh^2(\Psi),
  -a^2(t)\sinh^2(\Psi)\sin^2(\Theta))
\end{equation}
The energy--momentum tensor of matter is given by:
\begin{equation} 
{\cal T}^0_0  =  6\kappa_g\frac{\dot{a}^2 -1}{a^2};\qquad
 {\cal T}^1_1={\cal T}^2_2={\cal T}^3_3  =  2\kappa_g\frac{2\ddot{a}a +\dot{a}^2 -1}{a^2};
\end{equation}
While the gravitational energy--momentum tensor is given by:
\begin{equation} 
j^0_0  =  -6\kappa_g\frac{(\dot{a} -1)^2}{a^2}\qquad
 j^1_1=j^2_2=j^3_3  =  2\kappa_g\frac{(\dot{a} -1)^2}{a^2}
\end{equation}
But we do not reach the same conclusion: the total energy is not zero. This is
somewhat unexpected: if the total energy were zero in the three models, it
will agree quite well with the idea of originating from a state of zero total
energy. However this case is different and so, unless some other
interpretation is found, it raises the question of whether, although
mathematically possible, it is physically reasonable: why should universe
begin in a state of energy different from zero?. On the other hand, a flat
universe has the minimum matter density compatible with zero total energy.

\section{Last comments.}
We have already argued that equation (\ref{conserved1}) expresses the
conservation of energy--impulse, so let us write that equation as:
\begin{equation}
\Diamond_\nu S_{\mu}^{\;\;\nu}\equiv \nabla_\nu S_{\mu}^{\;\;\nu} - S_{\mu}^{\;\;\rho} T^\nu{}_{\nu\rho}=0
\end{equation}
We have also seen in equation (\ref{divergmat}) that Einstein's
equation implies:
\begin{equation}
\Diamond_\nu {\cal T}_{\mu}^{\;\;\nu} = - {\cal T}_{\sigma}^{\;\;\nu}T^\sigma{}_{\nu\mu}\label{nonconserv}
\end{equation}
Where ${\cal T}$ is the energy--momentum tensor of matter fields. Let us
suppose we are dealing with a perfect fluid in an isotropic homogeneous
universe. Let us consider for example the case of flat universe. The
energy--momentum tensor can be written in such case as:
\begin{equation}
{\cal T}_{\mu\nu}={\rm diag}\left(\rho,pa^2,pa^2,pa^2\right)
\end{equation}
Where $\rho=\rho(t)$ is the mass-energy density, $p=p(t)$ the pressure
and the $a^2(t)$ factors come from the metric.  The right hand of equation
(\ref{nonconserv}) can be easily computed and turns out to be:
\begin{equation}
\left(- 3\frac{\dot{a}}{a}p(t),0,0,0\right)
\end{equation}
The first thing which stands out is that if $p(t)\neq 0$ then in an
expanding universe, mass--energy of the material field (by itself) is
not conserved. We have seen such a behaviour when we considered the
radiation--filled universe before: a positive pressure means that the
gravitational field drains the positive energy from the ``material''
field. As a matter of fact we have also seen the case $p=0$ in the
dust-filled universe and there was no energy interchange. Let us turn
to the other possible case. Suppose there is some ``spontaneous matter
emission'' process, then if matter is created from the gravitational
field, pressure must be negative. Of course, we do not know what
exactly to put in the left hand, and the matter--emission process must
have very low probability of ocurrence because otherwise it would have
already been detected. Let us just put a small ``constant'' $\lambda$
in the left hand side of equation (\ref{nonconserv}), mainly because
we have no better guess, then we may write:
\begin{equation}
3p(t) =-\lambda \frac{a}{\dot{a}}
\end{equation}
So although the process may have very low probability of ocurrence,
the negative pressure increases with the expansion of the universe. It
may of course overcome the mass term in the equation which determines
the acceleration of the expansion of the universe (see for example
\cite{Wald84} pg. 97):
\begin{equation}
3\frac{\ddot{a}}{a}=-\frac{1}{4\kappa_g} \left[\rho(t) + 3p(t)\right]
\end{equation}
From that moment, positive acceleration sets in. We do not need to
have a cosmological constant to explain it. In fact we cannot consider
this $\lambda$ as a constant, it may depend on the strength of the
gravitational field, and also even in case we accept the possibility
of matter--emission processes, their rate must compensate the energy
absortion rate of the gravitational field from electromagnetic
radiation. We would need a quantum theory of gravitation to be able to
calculate $\lambda(t)$. However, we may get an idea of the order of
magnitude of $\lambda$ by considering zero the acceleration, taking
the Hubble constant $H_0=\dot{a}/a$ to be 70 (km/s)/Mpc, and taking
the density of the universe to be the critical one $\approx 2\times
10^{-26} \mbox{kg}/\mbox{m}^3$. We get $\lambda\approx 5\times
10^{-44}\mbox{kg}/(\mbox{m}^3\mbox{s})$. This is the order of
magnitude of the rate at which matter is created at the expense of the
gravitational field (of course, it says nothing about what sort of
particles are created). There is nothing to prevent gravitational
field from falling even further down in energy levels (as it is the
usual objection to negative energies), only that the rate is extremely
slow. Supposedly, a quantum theory of gravitation should be able to
explain this rate. Anyway it sets an experimental test for any such a
theory: see if within lowest order perturbation you can obtain
something similar.

\appendix
\section{The calculation of $Q$.}
First place we will need the explicit form of Einstein equations
written in terms of the torsion tensor and metric tensors. This has
been done quite a number of times, and we only put explicitly and
directly in terms of these tensor. After working out the calculations
implicit in equations (\ref{Euler}), Einstein's equations can be
written as:
{\setlength{\arraycolsep}{0.3mm}
\begin{eqnarray}\nonumber
 & &- \frac{1}{\kappa_g}{\cal T}_\sigma^{\;\;\nu}  =  a_1\left[2g^{\lambda \nu}\nabla_\sigma T^\alpha{}_{ \lambda \alpha} +\delta^\nu_\sigma (2\Lambda_0 - \Lambda_1)\right]  
 +  a_2\left[2\nabla_\gamma T^{\gamma \nu}{}_{\sigma} - 2\nabla_\gamma T^{\nu \gamma}{}_{\sigma} \right.\\ \nonumber
 & & - \left. 2g^{\lambda \nu}T^\alpha{}_{\beta \lambda}T^\beta{}_{\alpha \sigma}
  + 2g^{\lambda \nu}T^\gamma{}_{ \gamma \rho}T^\rho{}_{\sigma \lambda} -
 2g^{\lambda \rho}T^\gamma{}_{\gamma \rho}T^\nu{}_{\sigma \lambda} + \Lambda_2\delta^\nu_\sigma \right]  + \\
 & & + 
a_3\left[- 4T^{\rho \beta \nu}T_{\rho \beta \sigma}  -4T^\gamma{}_{ \gamma \rho}T_\sigma^{\;\;\;\nu \rho} +
  2T^{\nu \alpha \lambda}T_{\sigma \alpha \lambda}
 -  4g_{\sigma \tau}\nabla_\gamma T^{\tau \gamma \nu} + \Lambda_3\delta^\nu_\sigma \right]  \label{einstein}
\end{eqnarray}}
To accomplish the elimination of nontensorial elements in equation
(\ref{nontensorial}) let us do the needed calculations for each of the
three parts of the lagrangian density. For $\Lambda_1 = g^{\lambda
\mu}T^\alpha{}_{\alpha \lambda}T^\beta{}_{\beta \mu} $ one has
that:
\begin{equation}
h^a_{\;\;\nu ,\mu}\frac{\partial \Lambda_1}{\partial
    h^a_{\;\;\nu ,\gamma}}  -  \Lambda_1\delta^\gamma_\mu  =  
- 2\left( g^{\lambda\gamma}T^\alpha{}_{\alpha \lambda}\Gamma^\beta{}_{\beta \mu } 
  -  g^{\lambda \beta}T^\alpha{}_{\alpha \lambda}\Gamma^\gamma{}_{\beta \mu }\right)  -
\Lambda_1\delta^\gamma_\mu  
\end{equation}
And one may take:
\begin{equation}
 \frac{{}_1Q_\mu^{\;\;\gamma}}{\kappa_g} = 2\left( g^{\lambda\gamma}T^\alpha{}_{\alpha \lambda}T^\beta{}_{\beta \mu }  -
T^\alpha{}_{\alpha \lambda}T^{\gamma\lambda}{}_{\mu }\right)  -
\Lambda_1\delta^\gamma_\mu                         
\end{equation}
and
\begin{equation}
\frac{{}_1N_\mu^{\;\;\gamma}}{\kappa_g} = - 2\left( g^{\lambda\gamma}T^\alpha{}_{\alpha \lambda}\Gamma^\beta{}_{\mu \beta}  -
g^{\lambda \beta}T^\alpha{}_{\alpha \lambda}\Gamma^\gamma{}_{\mu \beta}\right)
\end{equation}
The subindex 1 of, for example, ${}_1N^\gamma_\mu $ just refers to
the fact that the term comes from $\Lambda_1$, it is not any sort of
spatial index. So after some work, which may involve using the
condition that the curvature of the Weitzenb\"ock connection is zero
for simplifying some expressions, one finds that:
\begin{eqnarray}\nonumber
\frac{1}{\kappa_g}\left[{}_1N_\mu^{\;\;\gamma}\Gamma^\nu{}_{\nu \gamma} + {}_1Q_\sigma^{\;\;\gamma}\Gamma^\sigma{}_{\mu\, \gamma} + \frac{\partial
  ({}_1N_\mu^{\;\;\gamma})}{\partial x^\gamma}\right] & = &\\
-\left(2g^{\lambda\nu}\nabla_\sigma T^\alpha{}_{\lambda\, \alpha}
+\delta^\nu_\sigma(2\Lambda_0 - \Lambda_1)\right)\Gamma^\sigma{}_{\mu \nu} & &
\end{eqnarray}
 For $\Lambda_2 = g^{\lambda \mu}T^\alpha{}_{\beta \lambda}T^\beta{}_{\alpha \mu}$ one gets:
\begin{equation}
h^a_{\;\;\nu ,\mu}\frac{\partial \Lambda_2}{\partial
    h^a_{\;\;\nu ,\gamma}} - \Lambda_2\delta^\gamma_\mu =
 -2\left( g^{\lambda\gamma}T^\alpha{}_{\beta \lambda}\Gamma^\beta{}_{\alpha \mu}  -
g^{\lambda\kappa}T^\gamma{}_{\beta \lambda}\Gamma^\beta{}_{\kappa \mu}\right)  - 
\Lambda_2\delta^\gamma_\mu 
\end{equation}
And then one may take:
\begin{equation}
\frac{{}_2Q_\mu^{\;\;\gamma}}{\kappa_g} = 2\left( g^{\lambda\gamma}T^\alpha{}_{\beta \lambda}T^\beta{}_{\alpha\, \mu}  -
T^\gamma{}_{\beta \lambda}T^{\beta \lambda}{}_{\; \mu}\right) -\Lambda_2\delta^\gamma_\mu
\end{equation}
and
\begin{equation}
\frac{{}_2N_\mu^{\;\;\gamma}}{\kappa_g} = -2\left( g^{\lambda\gamma}T^\alpha{}_{\beta\, \lambda}\Gamma^\beta{}_{\mu \alpha}  -
g^{\lambda\kappa}T^\gamma{}_{\beta \lambda}\Gamma^\beta{}_{\mu \kappa}\right)
\end{equation}
So after some second work, one finds that:
\begin{eqnarray}\nonumber
 &&\frac{1}{\kappa_g}\left[{}_2N_\mu^{\;\;\gamma}\Gamma^\nu{}_{\nu \gamma} + {}_2Q_\sigma^{\;\;\gamma}\Gamma^\sigma{}_{\mu\, \gamma} + \frac{\partial
  ({}_2N_\mu^{\;\;\gamma})}{\partial x^\gamma}\right]= 
 -2\Gamma^\sigma{}_{\mu \nu}\Big(\nabla_\gamma T^{\gamma \nu}{}_{\sigma } - \nabla_\gamma T^{\nu \gamma}{}_{\sigma} \\
 & & \left. - g^{\lambda\nu}T^\alpha{}_{\beta \lambda}T^\beta{}_{\alpha \sigma} + g^{\lambda\nu}T^\gamma{}_{\gamma \rho}T^\rho{}_{\sigma \lambda} - g^{\lambda\rho}T^\gamma{}_{\gamma \rho}T^\nu{}_{\sigma \lambda} + \Lambda_2\delta^\nu_\sigma\right)
\end{eqnarray}
Finally for $\Lambda_3 = g^{\lambda \mu}g^{\alpha \beta}g_{\rho \sigma}T^\rho{}_{\alpha \lambda}T^\sigma{}_{\beta \mu}$ one has:
\begin{equation}
h^a_{\;\;\nu ,\mu}\frac{\partial \Lambda_3}{\partial
    h^a_{\;\;\nu ,\gamma}} - \Lambda_3\delta^\gamma_\mu = 
 -4g^{\lambda\, \gamma}g^{\alpha \beta}g_{\rho \sigma}T^\rho{}_{\alpha \lambda}\Gamma^\sigma{}_{\beta \mu} - \Lambda_3\delta^\gamma_\mu  
\end{equation}
And then one may take:
\begin{equation}
\frac{{}_3Q_\mu^{\;\;\gamma}}{\kappa_g} =  4g^{\lambda\, \gamma}g^{\alpha \beta}g_{\rho \sigma}T^\rho{}_{\alpha \lambda}T^\sigma{}_{\beta \mu} - \Lambda_3\delta^\gamma_\mu
\end{equation}
and
\begin{equation}
\frac{{}_3N_\mu^{\;\;\gamma}}{\kappa_g} = -4g^{\lambda\, \gamma}g^{\alpha \beta}g_{\rho \sigma}T^\rho{}_{\alpha \lambda}\Gamma^\sigma{}_{\mu \beta} 
\end{equation}
So after some third work, one finds that:
\begin{eqnarray}\nonumber
\frac{1}{\kappa_g}\left[{}_3N_\mu^{\;\;\gamma}\Gamma^\nu{}_{\nu
  \gamma} + {}_3Q_\sigma^{\;\;\gamma}\Gamma^\sigma{}_{\mu\,
  \gamma} + \frac{\partial ({}_3N_\mu^{\;\;\gamma})}{\partial
  x^\gamma}\right]= \\ 
 -4\Gamma^\sigma{}_{\mu \nu}\left(-T^{\rho\beta\nu}T_{\rho\beta \sigma}
 - T^\gamma{}_{\gamma \rho}T_\sigma^{\;\;\nu \rho}
  +\frac{1}{2}T^{\nu\alpha \lambda}T_{\sigma\alpha \lambda} -
  \nabla_\gamma T_\sigma^{\;\; \gamma \nu} +
  \Lambda_3\delta^\nu_\sigma\right)
\end{eqnarray}
So taking into account Einstein's equation (\ref{einstein}),
one arrives to the conclusion that:
\begin{equation}
N_\mu^{\;\;\gamma}\Gamma^\nu{}_{\nu
  \gamma} +  Q_\sigma^{\;\;\gamma}\Gamma^\sigma{}_{\mu\,
  \gamma} + \frac{\partial (N_\mu^{\;\;\gamma})}{\partial
  x^\gamma}  = 
\Gamma^\sigma{}_{\mu \nu}{\cal T}_\sigma^{\;\;\nu}
\end{equation}

\end{document}